\documentclass[3p,times,procedia]{elsarticle}
\usepackage{nupha_ecrc}

\volume{00}

\firstpage{1}

\journalname{Nuclear Physics A}

\runauth{}

\jid{nupha}

\jnltitlelogo{Nuclear Physics A}

\usepackage{amssymb}

\usepackage[figuresright]{rotating}

\usepackage{graphicx}
\usepackage{dcolumn}
\usepackage{bm}
\usepackage{epstopdf}
\usepackage{mathrsfs}
\usepackage{amssymb,amsmath,amsfonts,latexsym}
\usepackage{nicefrac}
\usepackage[colorlinks=true,linktocpage=true,linkcolor=blue,citecolor=blue]{hyperref}
\usepackage[utf8]{inputenc}
\usepackage{lipsum}
\usepackage{nicefrac}
\usepackage{slashed}
\usepackage{mathtools}
\def\bs{\boldsymbol} 
\def\del{\partial}
\def\bdel{\bs\partial}
\newcommand{\eqn}[1]{Eq.~\eqref{#1}}

\long\def\comment#1{ }

\def\and{\qquad\text{and}\qquad} 
\def\el{\text{el}}

\def\BH{\text{BH}}

\def\mfp{\text{mfp}} 
\def\GLV{{\text{GLV-HT}}} 
\def\scatt{\text{scatt}}

\def\0{{\boldsymbol 0}}
\def\q{{\bm q}}

\def\x{{\boldsymbol x}}
\def\y{{\boldsymbol y}}

\def\qin{{\hat q_0}}

\def\HO{\text{HO}}
\def\pert{\text{pert}}

\def\Kc{{\cal K}}

\newcommand{\beq}{\begin{eqnarray}}
\newcommand{\eeq}{\end{eqnarray}}
\newcommand{\be}{\begin{eqnarray*}}
\newcommand{\ee}{\end{eqnarray*}}
\newcommand{\bal}{\begin{align}}
\newcommand{\eal}{\end{align}}
\newcommand{\rmd}{{\rm d}}

\newcommand{\rme}{{\rm e}}

\def\rmR{{\rm Re}}

\def\abar{{\rm \bar\alpha}}

\begin{document}

\begin{frontmatter}



\dochead{XXVIIIth International Conference on Ultrarelativistic Nucleus-Nucleus Collisions\\ (Quark Matter 2019)}

\title{Improved opacity expansion for medium-induced parton splitting}


\author{Yacine Mehtar-Tani}

\address{Physics Department and RIKEN BNL Research Center, Brookhaven National Laboratory, Upton, NY 11973, USA}

\begin{abstract}
We revisit the calculation of the medium-induced gluon radiative spectrum and propose a novel expansion scheme that encompasses the two known analytic limits: i) the high frequency regime dominated by a single hard scattering that corresponds to the leading order in the standard opacity expansion, ii) the low frequency regime that is dominated by multiple soft scatterings. Our approach is based on expanding around the harmonic oscillator instead of vacuum in the leading logarithmic approximation. We compute the first two orders in this improved opacity expansion and show that they account for the aforementioned limits. 
\end{abstract}


\end{frontmatter}


\section{Introduction}
\label{sec:intro}
The phenomenon of jet quenching, that is, the suppression of high pt jets (or hadrons) in ultra-relativistic heavy ion collisions (HIC) represents one of the crucial piece of evidence for the creation of the quark-gluon-plasma (QGP) in such collisions. It was first observed at RHIC \cite{Adcox:2001jp,Adler:2002tq} and confirmed later at the LHC \cite{Aad:2010bu,Chatrchyan:2011sx}. The observed strong suppression is due to the energy loss suffered by high energy partons, produced early in the collisions, while passing through the QGP. Although suppressed by an additional power of the coupling constant, radiative processes are enhanced in large media and hence, compete with elastic processes. 
Medium-induced radiative spectrum is therefore an important component for quantitative calculations of energy loss in the QGP and constitues the building block for medium-induced QCD cascade \cite{Jeon:2003gi,Blaizot:2013vha,Blaizot:2013hx,Schenke:2009gb,Zapp:2008gi}. However, owing to coherence effects multiple scattering during the quantum radiation process needs to be resummed to all order in the case of a dense medium  \cite{Baier:1996sk,Baier:1996kr,Zakharov:1996fv,Zakharov:1997uu,Arnold:2002ja}. This leads to the suppression of the gluon spectrum in the UV, for gluon energies larger than the Bethe-Heitler frequency. This is the QCD analog of the Laudau-Pomerantchuk-Migdal (LPM) in QED. 

There exist two known analytic limits: 1) Multiple soft scattering approximation, often referred to in the literature as the Baier-Dokshitzer-Mueller-Peigné-Schiff-Zakharov (BDMPS-Z) approximation. In this approximation, the in-medium mean-free-path is short, i.e.,  $\ell_\mfp \ll L$, hence, high density effects must be resummed to all orders. During the gluon formation time $t_f\sim \omega/k_\perp^2$ the transverse momentum accumulated via diffusion is given by $k_\perp^2\sim t_f  \hat q$, where $\hat q $ is the jet quenching transport coefficient, which to leading order in the coupling constant reads
\beq\label{eq:qhat-def-lo}
 \hat q(Q^2) \simeq  C_R\,\int_\q  \, q_\perp^2 \, \frac{\rmd \sigma_\el}{  \rmd^2 q_\perp} \simeq 4\pi \alpha_s^2 C_R n \ln\frac{Q^2}{\mu^2}. 
\eeq
where $\sigma_\el$ is the QCD elastic cross-section and $n$ the density of scattering centers. The logarithm $\ln\frac{Q^2}{\mu^2}$ is a result of the Coulomb tail which turns out to be the dominant contribution so long as the typical scale of the process $Q$ is much larger than the infrared cut-off scale $\mu$ that is related to the Debye mass.  In this regime, $t_f = \sqrt{\omega/\hat q } \ll L$ that corresponds to $ \omega \ll \omega_c =\hat q L^2/2$, the the problem reduces to that of a harminic oscillator and the solution writes
\beq\label{eq:bdmps}
 \omega\frac{\rmd I_{BDMPS}}{\rmd \omega} &=& 2 \abar  \ln \left|\cos\left(\frac{1-i}{2}\sqrt{\frac{2 \omega_c}{\omega}}\right)\right|\simeq  2 \abar
\,\,\sqrt{\frac{\omega_c}{2\omega}}   \quad\qquad\text{for}\quad \omega \ll \omega_c\\
\eeq
At large frequencies, $\omega >\omega_c$ the spectrum is dominated by a single hard scattering which is not accounted for in the BDMPS spectrum. The correct limit at large frequencies is given by the GLV spectrum 
{\it Single hard scattering approximation} (SHSA): In this approach the medium is assumed to be dilute, that is, $\ell_\mfp \gg L$. It was first discussed by Gyulassy, Levai and Vitev (GLV) \cite{Gyulassy:2000er}, and Wiedemann \cite{Wiedemann:2000za}.  A related approach, dubbed Higher-Twist (HT), involves an additional approximation, namely, that the gluon transverse momentum is much larger than the Debye mass  \cite{Wang:2001ifa}. Hence, we find 
\beq\label{eq:glv}
\omega\frac{\rmd I_\GLV}{\rmd \omega} \simeq  \abar  \frac{\pi}{4} \left(\frac{\qin\,L^2}{2 \omega} \right) \quad\quad\text{for} \quad \omega\, \gg \,\omega_c,
\eeq
where $\hat q_0 = 4\pi \alpha_s^2 C_R n$. 
We propose in this work a novel analytic approach inspired by the Molière scattering theory \cite{Moliere}, where in place of expanding order by order in opacity, we split the elastic cross-section into a soft and hard part. The former is resumed to all orders accounting for density effects and the latter is treated as perturbation that naturally accounts for the hard part of the radiation cross-section. We compute the first two orders and show that we recover the BDMPS and GLV-HT limits \cite{Mehtar-Tani:2019tvy,Mehtar-Tani:2019ygg}.




\section{Improve opacity expansion}

The general expression for the medium-induced gluon spectrum off a high energy parton with energy $E$ \cite{Baier:1996sk,Baier:1996kr,Zakharov:1997uu,Arnold:2002ja,Wiedemann:2000za} (see also \cite{Mehtar-Tani:2013pia,Blaizot:2015lma} for recent reviews) reads
\beq\label{eq:spectrum} 
\omega\frac{\rmd I}{\rmd\omega} &=&  \frac{\alpha_s C_R}{\omega^2} \, 2\rmR  \int_0^\infty \rmd t_2 \int_0^{t_2}\rmd t_1 \, \bdel_x \cdot \bdel_y\,  \Big[\Kc(\x,t_2|\y,t_1)- \Kc_0(\x,t_2|\y,t_1) \, \Big]_{\x=\y=\0}\,,
\eeq
where $\omega \ll E$. The Green's function $\Kc$ is solution of the Schr\"odinger equation:
\beq\label{eq:full-schordinger-0}
\left[ i \frac{\del}{\del t }  +\frac{\bdel^2}{2\omega} +i v(\x) \right] \Kc(\x,t|\y,t_1) =i \delta(\x-\y)\delta(t-t_1),
\eeq
where the imaginary potential $iv(\x)$ encodes the random transverse kicks that the two-body system made of the energetic parton and the radiated gluon  experiences while in the plasma. It is related to the in-medium elastic cross-section by the following Fourier transform,
$v (\x,t) = N_c\int_\q\, \, (\rmd\sigma_\text{el}/\rmd^2 \q)\left(1-\rme^{i\q\cdot\x}\right)$. The screening of the infrared (Coulomb) divergence depends on medium properties. In a thermal plasma, the HTL (Hard-Thermal-Loop) approximation yields \cite{Aurenche:2002pd} 
$ \rmd^2 \sigma_\text{el}/\rmd^2 \q \equiv g^2 m^2_DT/ (\q^2 (\q^2+m_D^2))$ 
where $m^2_D= (1+N_f/6) g^2 T^2$ is the QCD Debye mass squared, with $N_f$ the number of active quark flavors and $T$ the plasma temperature.  In an other approach, static scattering centers are assumed \cite{Gyulassy:2000er}. This corresponds to the following elastic cross-section, $\rmd^2 \sigma_\text{el}/\rmd^2 \q\equiv g^4 n/(\q^2+\mu^2)^2$. These two models for the elastic cross-section differ only when the moment transfer is of order the cut-off $\mu\sim m_D$.  For dense enough media the relevant transverse momentum scales is much larger than $\mu$, .i.e., $1/x_\perp^{2}\sim Q^2 \sim \mu^2 N_\scatt \sim g^4 n L \gg \mu$, as a result the sensitivity of the gluon spectrum to the infrared is reduced due to the weak dependence of the large Coulomb logarithm in $\mu$:
\beq\label{eq:dipole-log}
v(\x,t)  \equiv \frac{1}{4} \hat q(\x^2,t)\, \x^2 \simeq \frac{g^4}{16\pi} N_c\,n(t) \, \x^2 \, \ln\frac{1}{\mu^2\,\x^2 },
\eeq
where we have used the leading logarithmic definition of the transport coefficient \eqn{eq:qhat-def-lo}.  

The standard opacity expansion amounts to calculating the spectrum order by order in the potential $v$. In our approach on the other hand we expand around the harmonic oscillator, such that the leading order reproduces the BDMPS approximation \eqn{eq:bdmps} and the next-to-leading order (NLO) accounts for the single hard scattering limit \eqn{eq:glv}. For that purpose, we recall that $x_\perp \sim Q$, where the transverse scale $Q$ is related to the typical transverse momentum broadening of the gluon during the radiation process and split the scattering potential as follows:
\beq
v(t,\x)=\frac{1}{4}\qin(t)\, \x^2 \left(\ln\frac{Q^2}{\mu^2} +\ln\frac{1}{\x^2 Q^2} \right)\equiv v_\HO(t,\x)+v_\pert(t,\x),
\eeq
and treat $v_\pert(t,\x)$ as a perturbation assuming 
\beq 
\ln\frac{Q^2}{\mu^2}  \, \gg\, \ln\frac{1}{\x^2 Q^2}. 
\eeq
\section{Results}
\begin{figure}[]
\center
\includegraphics[width=6.5cm]{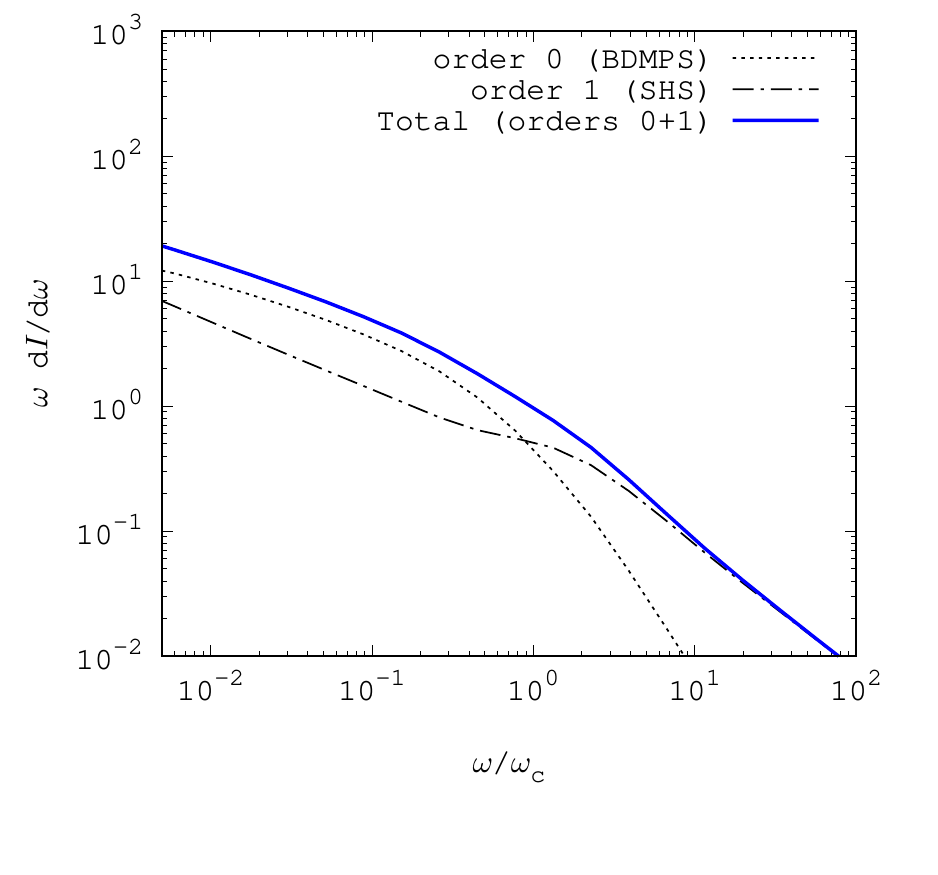}\includegraphics[width=6.5cm]{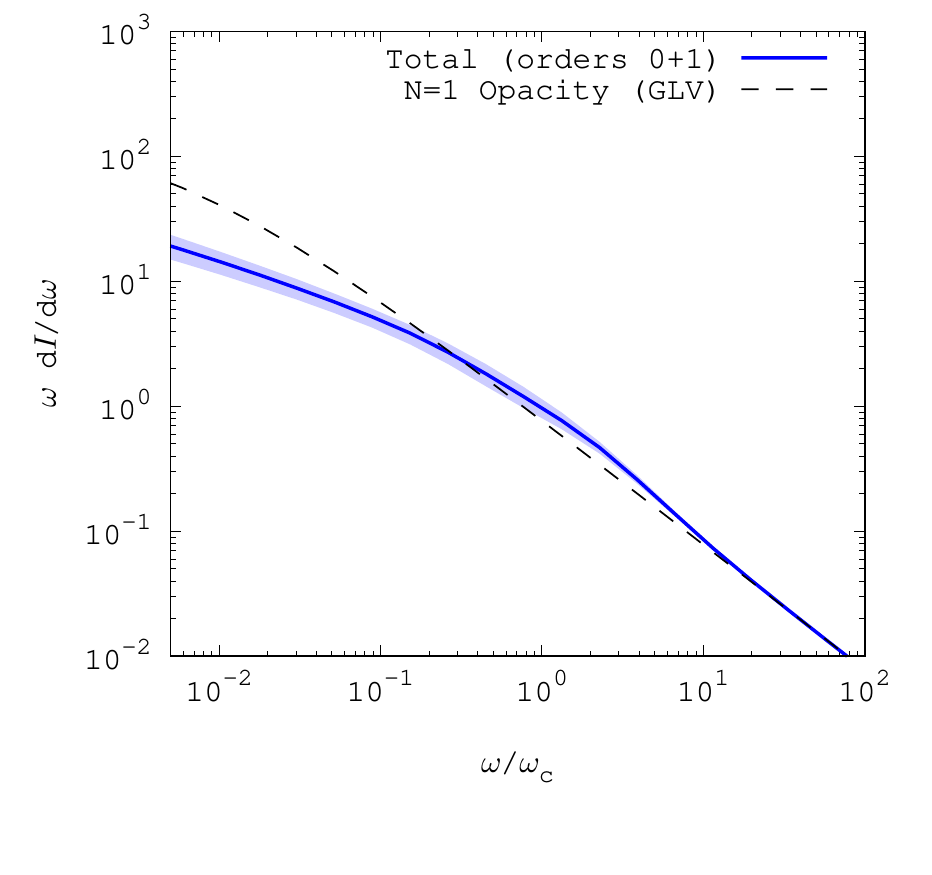}
\vspace{-0.5cm}
\caption{Left:The medium-induced spectrum to NLO in the expansion around the harmonic oscillator. The leading order (order 0) corresponds to the the BDMPS spectrum (dotted) and the first correction (order 1) accounts for to single hard scattering regime (dotted-dashed). The total spectrum is represented by a blue solid line. We used the following set of parameters: $\qin=0.1$ GeV$^{3}$, $L=4$ fm, $\mu=0.3$ GeV and we have set the overall pre-factor $\abar=1$. This corresponds to $\hat q \sim \qin  \ln (\qin L / \mu^2) = 1.5$ GeV$^2$/fm, $\omega_\BH \simeq 0.08$ GeV and $\omega_c \equiv \qin L^2 =40$ GeV. Right:The medium-induced spectrum to NLO in the expansion around the harmonic oscillator (BDMPS) (blue solid line). The blue band represents the variation of the IR transverse scale between $\mu^2/2$ and $2 \mu^2$, $\mu=0.3$ GeV being the central value. The GLV spectrum (dashed) is also plotted for comparison.}\label{fig:spectrum-new}
\end{figure}

To NLO accuracy we find \cite{Mehtar-Tani:2019tvy}
\beq\label{eq:total-spectrum}
 \omega\frac{\rmd I}{\rmd \omega }\simeq 2\,\abar \, \ln |\cos(\Omega L)|  + \frac{1}{2}\,\abar\,\qin \, \rmR  \int_0^L \rmd s \,  
\frac{1}{k^2(s)} \left[ \ln\frac{k^2(s)}{Q^2} +\gamma_E\right].
\eeq
where $k^2(s)= i  \frac{\omega \Omega}{2} (\cot(\Omega s)-  \tan(\Omega(L-s)))$,
and $\Omega=(1-i)/2\sqrt{\hat q(Q^2)/\omega}$. The separation scale is chosen to be the typical transverse momentum scale in the HO approximation
\beq
Q^2 \simeq \sqrt{ \omega \hat q (Q^2)} \simeq \sqrt{ \omega \qin\ln(\sqrt{\omega \qin }/\mu^2)}  \
\eeq
The first term is the standard BDMPS-Z result with however a $Q^2$ dependences that is cancelled partially by the second term. The latter, encompasses the correct UV behavior of the GLV spectrum that is $\omega^{-1}$ as shown in the Figure \ref{fig:spectrum-new}. 

The results are plotted in Figure \ref{fig:spectrum-new}, where we see that the BDMPS (LO) spectrum dominates in the infrared while in the UV the NLO takes over. The sensitivity to the infrared scale $\mu$ is also shown. This approach provides a simple analytic formula for medium-induced gluon spectrum, so long as $\omega >\omega_\BH$, that can be used in phenomenological studies of jet quenching with reduced theoretical uncertainties related to the use of the BDMPS or GLV-HT approximations. In particular, the obtained spectrum can be implemented in future Monte Carlo event generators of in-medium parton cascades.

\section*{Acknowledgements} 
This work is supported by the U.S. Department of Energy, Office of Science, Office of Nuclear Physics, under contract No. DE- SC0012704, the RHIC Physics Fellow Program of the RIKEN BNL Research Center
and in part by Laboratory Directed Research and Development (LDRD) funds from Brookhaven Science Associates. 
\bibliographystyle{elsarticle-num}
\bibliography{<your-bib-database>}









\end{document}